\documentclass[aps,prc,twocolumn,showpacs,amsmath,amssymb,nofootinbib]{revtex4}
\usepackage{epsfig,colordvi}

\usepackage{graphicx}
\usepackage{dcolumn}
\usepackage{bm}
\usepackage{epsfig}
\usepackage{fancyhdr}
\usepackage{array}
\usepackage{amsmath}
\usepackage{color}

\usepackage{graphicx}

\usepackage[bookmarks=true]{hyperref}
\usepackage{color}

  \hypersetup{
   colorlinks=true,
   pdfborder={0 0 0},
   citecolor=blue,
   linkcolor=blue,
   urlcolor=blue
 }

\def\be{\begin{equation}}

\def\ee{\end{equation}}
\def\barr{\begin{array}}
\def\earr{\end{array}}

\def\nn8{\nonumber\\[2pt]}

\def\ed{\end{document}}

\begin{document}

\title{Shell model results for $T=1$ and $T=0$ bands in $^{66}$As}

\author{Praveen C. Srivastava$^{1}$\footnote{pcsrifph@iitr.ac.in}, 
R. Sahu$^{2}$\footnote{rankasahu@rediffmail.com} and 
V.K.B. Kota$^3$\footnote{vkbkota@prl.res.in}}

\address{$^{1}$Department of Physics, Indian Institute of Technology Roorkee,
Roorkee 247 667, India}

\address{$^{2}$ National Institute of Science and Technology,
Palur Hills, Berhampur-761008, Odisha, India}

\address{$^{3}$Physical Research Laboratory, Ahmedabad 380 009, India}

\date{\hfill \today}

\date{\hfill \today}

\begin{abstract}

Results of a comprehensive shell model (SM) analyses, within the full
$f_{5/2}pg_{9/2}$ model space, of the recently available experimental data [P.
Ruotsalainen et al., Phy. Rec. C {\bf 88}, 024320 (2013)] with four $T=0$ bands
and one $T=1$ band in the odd-odd $N=Z$ nucleus $^{66}$As are presented.  The
calculations are performed using jj44b effective interaction developed
recently by B.A. Brown and A.F. Lisetskiy for this  model space.  For the lowest
two $T=0$ bands and the $T=1$ band, the results  are in reasonable agreement
with experimental data and deformed shell model is used to identify their
intrinsic structure. For the $T=1$ band, structural change at  $8^+$ is
predicted. For the third $9^+$ band with $T=0$, the shell model $B(E2)$ values
and quadrupole moments (in addition to energies) are consistent with the
interpretation in terms of  aligned isoscalar $np$ pair in $g_{9/2}$ orbit
coupled to the $^{64}$Ge ground band. Similarly, the $9^+$ level of band 4 and a
close lying $5^+$ level are found to be isomeric states in the analysis.
Finally, energies of the  band 5 members calculated using shell model with both
positive and negative parity show that the observed levels are most likely
negative parity levels. The SM results with jj44b are also compared  with the
results obtained using JUN45 interaction. 

\end{abstract}


\maketitle

\section{Introduction}

There has been considerable interest in investigating the structure of the
nuclei in the mass region $A = 60-100$ and in particular even-even and odd-odd 
$N = Z$ nuclei. The $N=Z$ nuclei in this mass region lie near the proton
drip-line. The even-even $N=Z$ nuclei in this region exhibit rapid changes in
nuclear shape  and nuclear structure with changing nucleon number. For example,
$^{64}$Ge exhibits $\gamma$-soft structure \cite{Sta-07}, $^{68}$Se exhibits
oblate shape in the ground state \cite{68se}, $^{72}$Kr \cite{72kr1,72kr2,72kr3}
exhibits shape coexistence, $^{76}$Sr and $^{80}$Zr have large ground state
deformations \cite{76Sr,Lister-87} and so on.  Recently, evidence for a
spin-aligned neutron-proton isoscalar paired phase  has been reported from the
level structure of $^{92}$Pd \cite{cederwall}.  Similarly, $\beta$ decay of the $T=1$
($J^\pi=0^+$) and $T=0$ ($J^\pi=9^+$) isomers in $^{70}$Br  have been recently
reported in Ref. \cite{70Br_PRC95}. Also many even-even $N = Z$ nuclei in this
region are of astrophysical interest since they are waiting point nuclei for
rp-process  nucleosynthesis \cite{Fuller}. Though even-even $N=Z$ nuclei are
important and interesting, in the recent years there has been special focus on
odd-odd $N=Z$ nuclei as these nuclei are expected to  give new insights into
neutron-proton ($np$) correlations that are hitherto  unknown.   

With the development of radioactive ion beam  facilities and large detector
arrays, new experimental results for the energy spectra of $N=Z$ odd-odd nuclei
starting from $^{62}$Ga to $^{86}$Tc  are now available in this region; see 
\cite{ks-book}  and references cited therein. These studies have opened up
challenges in describing and predicting spectroscopic  properties of these
nuclei.  In comparison to lighter $fp$ shell nuclei, production cross section in
fusion-evaporation reactions become very small for nuclei north of $^{56}$Ni. 
The recent development of the recoil-$\beta$-tagging technique provides a tool
to study medium-mass nuclei around the $N=Z$ line. Using this several  $T=0$ and
$T=1$  levels in the odd-odd $N=Z$ $^{62}$Ga nucleus were identified in
\cite{Ga62}.  Recently, we have been successful in using the  shell model (SM)
and deformed shell model (DSM) with jj44b interaction (due to Brown and
Lisetskiy \cite{jj44b}) to study comprehensively the $T=0$ and the $T=1$
bands/levels in $^{62}$Ga  \cite{Ga62-smdsm}. Turning to the next odd-odd $N=Z$
nucleus $^{66}$As, recently large number of excited states of $^{66}$As were
populated using $^{40}$Ca($^{28}$Si,$pn$)$^{66}$As  fusion-evaporation reaction
at beam energies of 75 and 83 MeV by Jyv\"{a}skyl\"{a} group \cite{66as}. Also
in this experiment half-lives and ordering of the two known isomeric states
($5^+$ at  1354 keV and $9^+$ at 3021 keV) have been determined with improved
accuracy.  Besides this, the experimental data has resulted in identifying  five
bands in this nucleus.  

Theoretical studies for low-lying  and low spin $T=0$ and $T=1$ states of
$^{66}$As were carried out using shell model (SM) \cite{66as},  deformed shell
model (DSM) \cite{sk1} and  IBM-4 \cite{ibm-4} in the past.  The above SM
calculation was performed using the JUN45 interaction \cite{jun45} and
similarly, the DSM calculations used a much older interaction due to
Madrid-Strasbourg group \cite{No-01}. Similarly, IBM-4 uses a Hamiltonian
obtained using a mapping procedure that  employs the underlying $SU(4)$ algebra.
The aim of the present study is to explain, more comprehensively, the recent
experimental data for $^{66}$As using shell model (SM)  and to bring out  the
structure of all the five bands observed in this nucleus by employing the same
jj44b interaction we used before for $^{62}$Ga. In addition, DSM
\cite{ks-book} is also used to bring out the structure of the intrinsic states
generating bands 1-3 of this nucleus. Some of the results presented in this
paper are first reported in \cite{rila-2016,pcs(R)}. As discussed ahead, the
shell model results for band 4 (high-spin states) are explained using Cranked
Nilsson-Strutinsky  (CNS) calculations \cite{pcs(R)}. Now we will give a preview.

Section II gives the model space, effective  interaction and other calculation
details. The structure of each of the five  observed bands are discussed using
SM in Sections III.A to III.E. Results are also compared with those from other
calculations where available. Finally, concluding remarks are drawn in  
Sect. IV.

\section{Method of calculations}

In the SM calculation, $^{56}$Ni is taken as the inert core  
with  the spherical orbits $1p_{3/2}$, $0f_{5/2}$, $1p_{1/2}$ and $0g_{9/2}$ 
forming the basis space. The jj44b interaction developed by  Brown and
Lisetskiy \cite{jj44b} has been used in both the calculations.  This interaction
was  developed by fitting with 600 binding energies and excitation energies of
nuclei with $Z = 28-30$ and $N = 48-50$ available in this region. Here, 30
linear combinations of $JT$ coupled two-body matrix elements (TBME) are varied
giving the rms deviation of about 250 keV from  experimental data. The single
particle energies (spe)  are taken to be  $-9.6566$, $-9.2859$, $-8.2695$ and $-5.8944$
MeV for the $p_{3/2}$, $f_{5/2}$, $p_{1/2}$ and $g_{9/2}$ orbits, respectively
\cite{jj44b}. Shell model calculations are performed using the shell model
code  Antoine \cite{antoine}. The maximum  matrix dimension in $M$-scheme is for
$0^+$ states ( $>$ $21$ million). 

\begin{figure} 
\includegraphics[width=8.5cm]{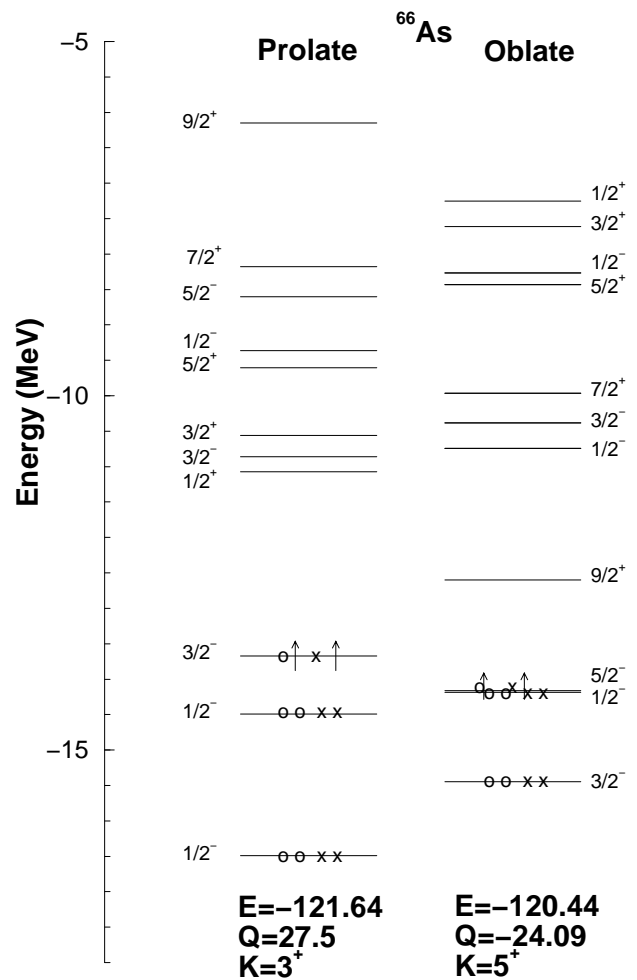} 
\caption{HF single-particle spectra for $^{66}$As corresponding to the lowest
prolate and oblate configurations. In the figure circles represent protons and
crosses represent neutrons. The HF energy (E) in MeV, mass quadrupole moment (Q)
in units of the square of the oscillator length parameter and the total azimuthal
quantum number K of the lowest intrinsic states are given in the figure.}
\label{fig-3pt12}
\end{figure}

In DSM, using the same set of single particle (sp) orbitals, spe and TBME,
as used in the SM calculation, the lowest energy intrinsic states
(prolate and oblate) for $^{66}$As are obtained by solving the Hartree-Fock
(HF) single particle equation self-consistently assuming axial symmetry. Excited
intrinsic configurations are obtained by making particle-hole excitations over
the lowest intrinsic state. Good angular momentum states are projected from
different intrinsic states and after isospin projection and  orthonormalization,
band mixing  calculations are performed as described in \cite{ks-book}.
Fig.~\ref{fig-3pt12} gives the HF sp spectrum for both prolate
and oblate solutions. The total isospin for the lowest configurations shown in
Fig.~\ref{fig-3pt12} is clearly $T=0$. By making particle-hole excitations for 
the six nucleons out side the lowest $k$ orbit, we have considered 114
configurations as given in \cite{rila-2016} and they generate forty four $T=0$
and fifty $T=1$ deformed configurations.

\begin{figure*}
\includegraphics[width=16.5cm]{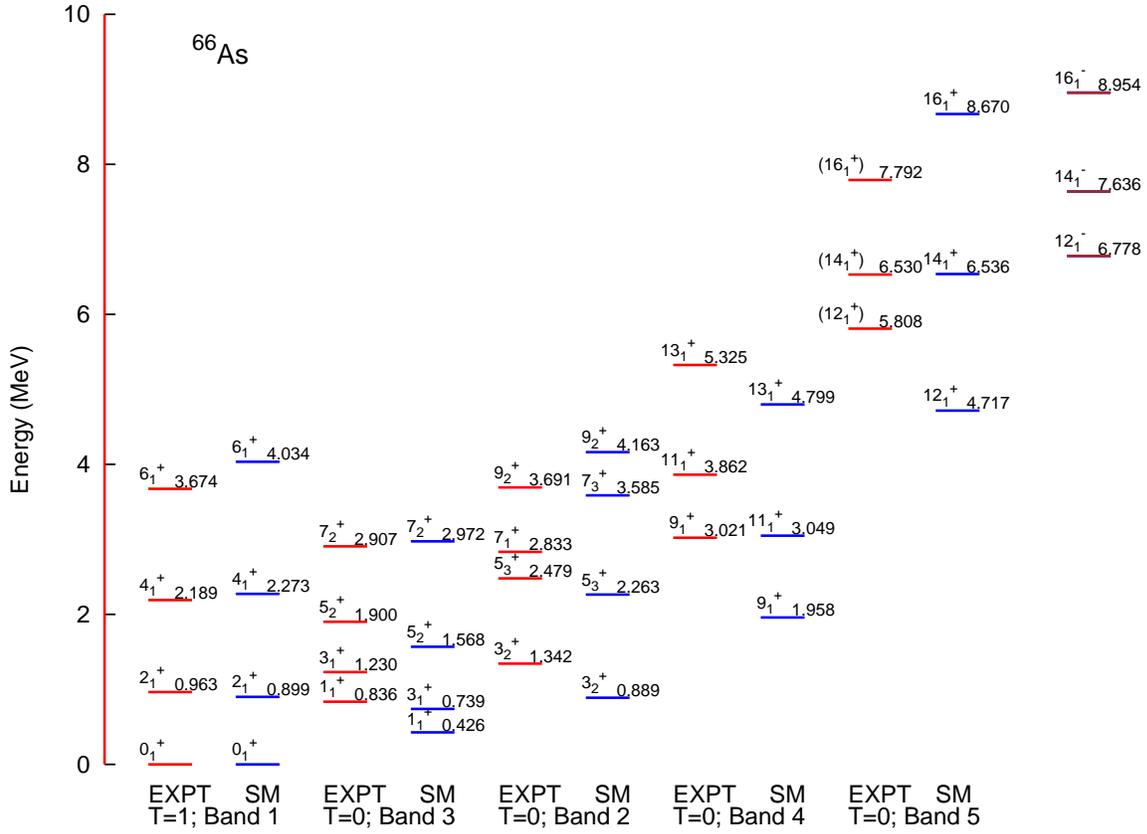}

\caption{Comparison of shell model results with experimental data for
different bands with jj44b interaction. The band numbers in the figure are as
per the convention used in the experimental paper \cite{66as}.
The jj44b interaction predicting isomeric $5^+$ state at 985 keV, while experimental value 
is 1354 keV. We have not shown this state in the present figure.
}
\label{bandsm}
\end{figure*}

\begin{figure*}
\includegraphics[width=12.5cm]{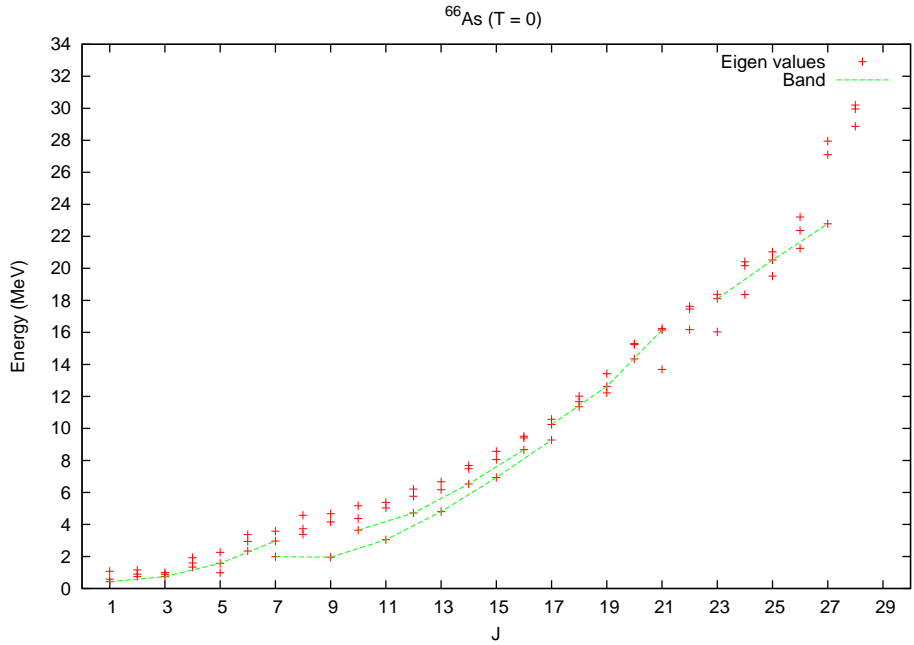}
\caption{Shell model predictions for different bands in $^{66}$As.}
\label{bandsmocc}
\end{figure*}
\begin{figure}
\includegraphics[width=6cm]{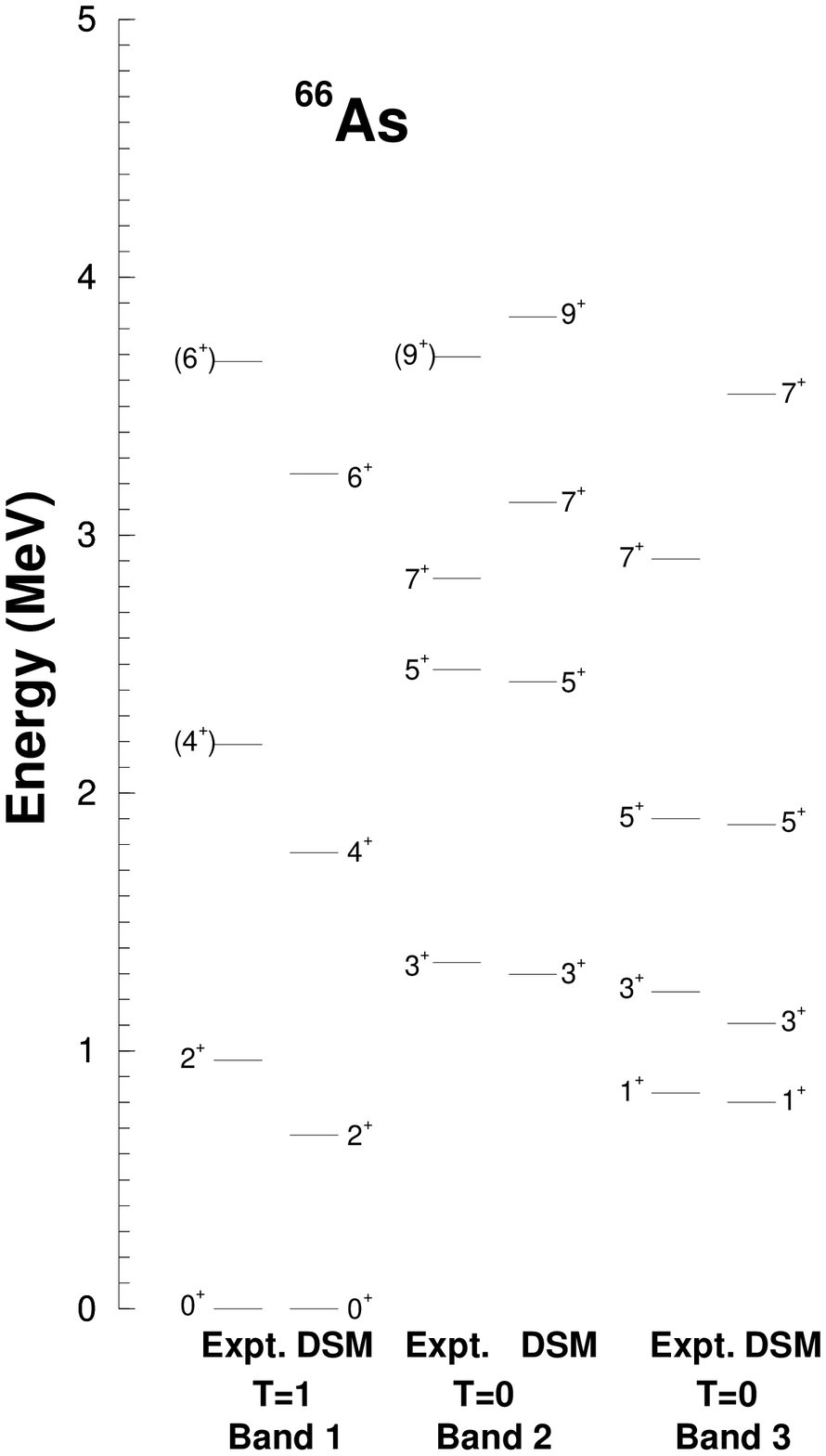}

\caption{ Comparison of deformed shell model results with experimental data 
for different bands with jj44b interaction.
}  
\label{66as_dsm}
\end{figure}

\section{Results and discussions}

In  Fig.~\ref{bandsm}, the SM results for $T=0$ and $T=1$ bands are compared
with experiment. In SM many levels are calculated (for each $J^\pi$) and they
are classified into different bands on the basis of dominant $E2$ transitions  
between them. As shown in Fig.  ~\ref{bandsmocc},  for identifying band
structures, we have connected by lines the states with strong transition matrix
elements between them and with similar dominant configuration in the wave
functions. Occupancies of the orbits for the levels in various bands are shown
in Table I. In DSM, calculated levels are classified into bands based on the
dominant intrinsic configuration \cite{ks-book} and the results are shown in
Fig. ~\ref{66as_dsm} for the lowest 3 bands. The agreements with experiment are
reasonable for SM for all the five bands and for the lowest three bands for
DSM. We will discuss below the structure of these bands in detail and also
compare the results with those obtained using JUN45 interaction reported in
\cite{66as}.

\begin{table}
\begin{center}
\caption{Shell model occupancies for $T=0$ and $T=1$ bands with jj44b/JUN45 interactions.
We have shown occupancies of band 1, band 2, band 3 and band 4 with JUN45 for comparison.}
\vspace{0.2cm}
\label{tab:table1}
\begin{tabular}{ c  c } \hline 
\hline
     &   nucleon occupation numbers      \\ 
        &   $n_{lj}^\pi$ =  $n_{lj}^\nu$ ($p_{3/2}$,$f_{5/2}$, $p_{1/2}$,
	$g_{9/2}$)   \\
\hline
T=1          &       Band 1   \\
\hline
$0^+$          &    1.97/2.49~ 1.81/1.49~ 0.56/0.66~ 0.66/0.36     \\
$2^+$         &    1.89/2.35~ 1.91/1.60~ 0.56/0.69~ 0.64/0.35      \\
$4^+$         &    1.83/2.42~ 2.00/1.64~ 0.53/0.62~ 0.64/0.32      \\
$6^+$          &    1.79/2.30~ 2.09/1.82~ 0.48/0.57~ 0.64/0.30      \\
$8^+$          &    1.45/2.60~ 1.92/1.76~ 0.47/0.42~ 1.16/0.21      \\

\hline
T=0            &       Band 2   \\
$3^+$          &    2.12/2.49~ 1.87/1.64~ 0.48/0.59~ 0.53/0.27      \\
$5^+$          &    2.15/2.49~ 1.81/1.69~ 0.52/0.59~ 0.52/0.22      \\
$7^+$         &    1.34/1.59~ 2.00/1.86~ 0.37/0.47~ 1.28/1.07      \\
$9^+$           &    1.50/2.08~ 1.77/2.19~ 0.49/0.54~ 1.24/0.19      \\

\hline
 T=0           &       Band 3  \\
$1^+$        &    1.96/2.47~ 2.03/1.60~ 0.53/0.67~ 0.48/0.25      \\
$3^+$          &    1.91/2.46~ 2.11/1.68~ 0.51/0.61~ 0.48/0.25      \\
$5^+$         &    2.52/2.32~ 1.26/1.85~ 0.83/0.57~ 0.39/0.25      \\
$7^+$        &    1.82/2.19~ 2.21/2.04~ 0.48/0.53~ 0.49/0.24      \\

\hline 
 T=0           &       Band 4   \\
$5^+$          &    2.52/2.74~ 1.26/1.23~ 0.83/0.85~ 0.39/0.18      \\
$9^+$         &    1.48/1.61~ 1.95/1.86~ 0.48/0.46~ 1.09/1.08      \\
$11^+$         &    1.52/1.68~ 1.88/1.76~ 0.51/0.49~ 1.08/1.07     \\
$13^+$          &    1.64/1.76~ 1.66/1.63~ 0.63/0.55~ 1.07/1.06     \\

\hline 
 T=0           &       Band 5 ($+$ve)  \\
$12^+$          &    1.54~ 1.89~ 0.47~ 1.09     \\
$14^+$           &    1.66~ 1.74~ 0.53~ 1.08      \\
$16^+$         &    1.95~ 1.36~ 0.63~ 1.06      \\

\hline 
 T=0        &       Band 5 ($-$ve)  \\
$12^-$          &    1.16~ 1.83~ 0.43~ 1.57     \\
$14^-$           &    1.20~ 1.74~ 0.45~ 1.59      \\
$16^-$         &    1.33~ 1.59~ 0.51~ 1.57     \\

\hline

\hline
\label{tocc}
\end{tabular}
\end{center}
\end{table}
\begin{table}
\begin{center}
\caption{ Comparison of $B(E2)$'s values for different bands with jj44b and JUN45 interactions. Results
are in $e^2fm^4$ with $e_p=1.5$e and $e_n=0.5$e.}
\vspace{0.2cm}
\label{tab:table2}
\begin{tabular}{ c  c  c} \hline 
\hline 
Transitions     &   jj44b    &     JUN45     \\ 
\hline
  &       Band 3   & \\
  
$7^+$ $\rightarrow$   $5^+$      &   416.12   &   374.91    \\ 
$5^+$ $\rightarrow$   $3^+$      &   375.98    &   333.81   \\ 
$3^+$ $\rightarrow$   $1^+$       &   296.63   &   204.00   \\ 

\hline
  &       Band 2   & \\
  
$9^+$ $\rightarrow$   $7^+$       &    277.62   &   0.27    \\ 
$7^+$ $\rightarrow$   $5^+$       &    0.01   &   0.000  \\ 
$5^+$ $\rightarrow$   $3^+$       &    38.33   &   55.56    \\ 

\hline
  &       Band 4   & \\
  
$13^+$ $\rightarrow$   $11^+$      &   282.85     &   266.66   \\ 
$11^+$ $\rightarrow$   $9^+$        &   326.96   &   306.06    \\ 

\hline

\hline
\label{tocc2}
\end{tabular}
\end{center}
\end{table}

\begin{table}[h]
\begin{center}
\caption{$B(M1)$ values in $\mu_N^2$. Here $g_s=g_{free}$ used in the SM calculations.}
\label{tab:table3}
\begin{tabular}{ c | c   } 
\hline \hline

$I_f^+ \rightarrow I_i^+$   & jj44b \\ \hline

T=1 $\rightarrow$ T=0       &              \\
\hline
$0_1^+ \rightarrow 1_1^+$     &  0.16         \\

$2_1^+ \rightarrow 1_1^+$   & 0.073       \\

$2_1^+ \rightarrow 3_2^+$     & 0.051        \\

$2_1^+ \rightarrow 3_1^+$     & 0.023        \\

$4_1^+ \rightarrow 5_2^+$    & 0.0048        \\

$4_1^+ \rightarrow 5_1^+$    & 0.0052         \\

$6_1^+ \rightarrow 7_2^+$    & 0.025        \\

$6_1^+ \rightarrow 7_1^+$    & 0.0004              \\

\hline
\hline

\end{tabular}
\end{center}
\end{table}
\subsection{Band \# 1 $(T=1)$}

The $T=1$ band (band \#1 in Fig. ~\ref{bandsm}) with $0^+$, $2^+$, $4^+$ and 
$6^+$ is well described by shell model  and it is the isobaric analogue of the
lowest $T=1$ band in $^{66}$Ge; see \cite{nndc}; all the levels are
quantitatively reproduced.  The DSM calculated  $T=1$ band (Fig.
~\ref{66as_dsm}) also agrees  reasonably well  with experiment.  Except for the
$2^+ \rightarrow 0^+$ separation (it is lower compared to experiment by $\sim
300$ keV),  the relative spacing of all other levels are reasonably reproduced.
 This implies that DSM predicts this band to be more collective and hence more 
compressed compared to data and SM results.
The $T=1$ levels up to $J=6^+$ mainly originate from the lowest $T=1$ intrinsic
state generated by the antisymmetric combination of the configurations
$(1/2^-)_1^{2p,2n}\,(1/2^-)_2^{2p,2n}  (3/2^- \uparrow)_1^{p} (3/2^-
\downarrow)_1^{n}$ and $(1/2^-)_1^{2p,2n}\,(1/2^-)_2^{2p,2n}  (3/2^-
\uparrow)_1^{n} (3/2^- \downarrow)_1^{p}$. Hence, there is no change in the
collectivity up to $J=6^+$. The shell model (SM) as well as the DSM predicts the
$B(E2)$ values for the transition $8^+ \rightarrow 6^+$ to be very small (their
energies are 5.124 MeV in SM and 3.973 in DSM).  For example, the $B(E2)$
ratios  $B(E2,I \rightarrow I-2)$ $/B(E2,I-2\rightarrow I-4)$ with I=4, 6, 8 are
1.22, 0.97 and 0.001 in DSM. The corresponding ratios for shell model are 1.29,
1.09 and 0.001.   The occupancy of the $1g_{9/2}$ orbit as seen in Table 
~\ref{tocc}  does not change much up to spin $T=1, J=6^+$ and is about 0.64 for
both protons and neutrons.  However, as we go to $T=1, 8_1^+$ level, there is a
dramatic change in the $g_{9/2}$  occupancy which is 1.16 in SM. Thus, shell
model predicts the structure of the $T=1, 8_1^+$ level to be quite different
from that of the other $T=1$ levels lying below. As a result, the $B(E2)$
transition probability from $T=1, 8^+$ to $T=1, 6^+$ is small. This is in
agreement with the conclusion drawn from the DSM calculation which predicts that
the structure of the $T=1, 8_1^+$ level to be quite different from that of the
$T=1,6^+_1$ level. This level originates from the $T=1$ projected intrinsic
state in which three protons and three neutrons are distributed in six single
particle orbitals. This configuration has a proton and a neutron in $g_{9/2}$
orbit just the occupancy given by SM. With the structure of the $T=1, 8^+_1$
level being quite different from the $T=1,  6^+_1$, its $E2$ transition
probability  to $T=1, 6^+_1$ level is also small in DSM just as in SM. Let us
add that previous SM studies \cite{66as} of $^{66}$As were due to Honma et al
using JUN45 \cite{jun45} and Hasegawa et al using an extended pairing plus
quadrupole interaction \cite{Has-2005}. The SM with JUN45 gives \cite{66as} the 
$2^+$, $4^+$, $6^+$ and $8^+$ excitation  energies to be $0.967$, $2.222$,
$3.891$ and $5.604$ MeV respectively.    It is seen that the energies of the
$T=1$ band members ($2^+$, $4^+$, $6^+$) are slightly better with JUN45 compared to those with
jj44b. However, the structure of the $8^+$ predicted by JUN45 is quite different
with very low $g_{9/2}$ orbit occupancy.

\subsection{Band $\#$ 2 ($T=0$)}

Coming to the $T=0$ bands, band \#2   (as classified in \cite{66as}) is the
first $T=0$ band and it is a $3^+$ band. It is seen from Fig.   ~\ref{bandsm}
that the SM calculated spectrum for band \#2 agrees reasonably well with 
experiment. Though data are not available, $B(E2)$ values within band \#2 are
calculated and SM gives for $5^+_3 \rightarrow 3^+_2$, $7^+_3 \rightarrow 5^+_3$
and $9^+_2 \rightarrow  7^+_3$ the values to be $38$, $0.010$ and $278$
$e^{2}fm^{4}$.  The effective charges used in this work are $e_p = 1.5e$ and
$e_n = 0.5e$.  The results of $B(E2)$ values for different bands with JUN45 and
jj44b are shown in the Table II. The small $E2$ transition between $7^+_3
\rightarrow 5^+_3$  is due to the structural change between these states. The
$3^+_2$ and $5^+_3$ wave functions is dominated by $0p-0h$ excitation more than
50 \%. On the other hand, the $7^+_3$ and $9^+_2$ states show (proton-neutron)
pair excitation to the $g_{9/2}$ orbit. The occupation of the $1p_{3/2}$ 
suddenly changes from $\sim$ 2.2 ($3_2^+$, $5_3^+$) to $\sim$ 1.3 ($7_3^+$,
$9_2^+$). This apparent structure change causes the small transition between
$7^+_3 \rightarrow 5^+_3$. Let us add that JUN45 interaction predicts
$3^+-5^+-7^+-9^+$ states at 995, 2154, 2652 and 4677 keV, respectively. While
corresponding jj44b interaction results are 889, 2263, 3585, and 4163 keV,
respectively. Also, as seen from the occupancies in Table I and $B(E2)$ for
$9/2^+$ to $7/2^+$ given in Table II, the structure of the $9^+$ level in
this band predicted by JUN45 is quite different from the structure predicted by
jj44b interaction.
Finally, Fig.~\ref{66as_dsm} shows the band \#2 levels from DSM.
It is seen that these levels have admixtures from many intrinsic states (except
for the $3^+$ level) and therefore DSM do not give this to be a proper
collective band. Thus, measuring
B(E2)'s in future for band \#2 members is important.
We have also shown the $M1$ transitions between $T=1$ to $T=0$ states in Table III.
These values are very small, although the $B(M1\, 6^+_1 \rightarrow  7^+_2)$ estimated to be
$\sim$ 1 $\mu_N^2$ in Ref. \cite{66as}. The JUN45 interaction result is also small \cite{66as}.

\subsection{Band $\#$ 3 ($T=0$)}

Band \#3 with  $T=0$ consists of $1^+$, $3^+$, $5^+$ and $7^+$ levels. All these
levels except $7^+$ level are found to have similar structure both in SM  and 
DSM. They mainly originate, as seen from DSM, from the symmetric  combination of
the configurations $(1/2^-)_1^{2p,2n}\,(1/2^-)_2^{2p,2n}  (3/2^- \uparrow)_1^{p}
(3/2^- \downarrow)_1^{n}$ and $(1/2^-)_1^{2p,2n}\,(1/2^-)_2^{2p,2n}  (3/2^-
\uparrow)_1^{n} (3/2^- \downarrow)_1^{p}$ with admixtures from the lowest
intrinsic state shown in Fig. 1 and several other intrinsic states. Also, for
this band, the JUN45 predict $1^+-3^+-5^+-7^+$ states at 611, 871, 1520, and
2750 keV, respectively while jj44b predicts at 426, 739, 1568 and 2972 keV,
respectively. Experimentally, no measurements have been made regarding the
$B(E2)$ values among the levels of this band. In SM, the calculated $B(E2)$'s
for   $3^+_1 \rightarrow 1^+_1$, $5^+_2 \rightarrow 3^+_1$ and $7^+_2
\rightarrow  5^+_2$ are $297$, $376$ and $416$ $e^{2}fm^{4}$.  In addition, the
$B(M1)$ value for the transition $1^+_1$ to ground $0^+$ is $0.0832~ \mu_N^2$
and measurement of this $B(M1)$ will give a good test of the structure of band
\#3. The $B(E2)$ for the decay of $3^+$ of band \#2 to  $1^+$ of band \#3 will
also give a good test of the structure of this band and the shell model value
for this transition is $0.70$ $e^{2}fm^{4}$.

\begin{figure*}
\resizebox{1.0\textwidth}{!}{
\includegraphics{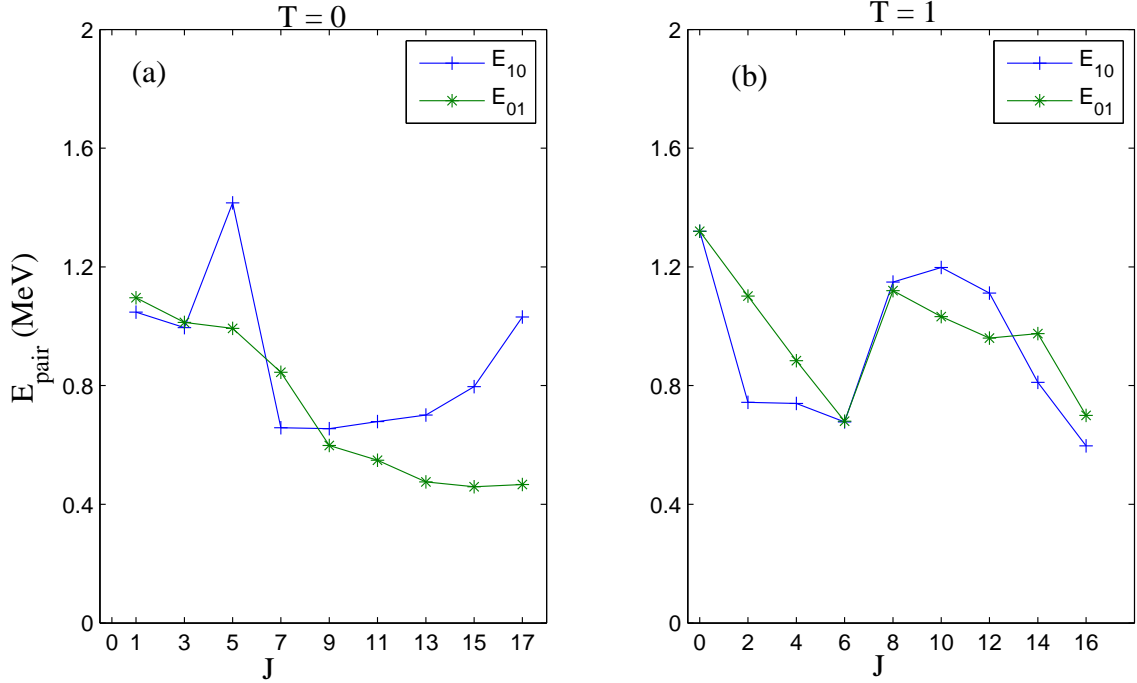} 
}
\caption{Pairing energies from shell model for (a) $T =0$ odd spin states and
(b) $T = 1$ even spin states. Here $E_{01}$ and $E_{10}$ are isovector and 
isoscalar pairing energies respectively.
}
\label{pairing}  
\end{figure*}

An important question that is being probed in the recent years is the energy
separation between the lowest $T=1$ and $T=0$ levels, i.e. between the ground
$0^+$ and the excited $1^+_1$ level shown in Fig. ~\ref{bandsm}.  To this end,
we have calculated the pairing energy following the procedure  discussed in ref.
\cite{povesplb}, i.e. by taking the energy difference of  states calculated with
the full Hamiltonian jj44b and the Hamiltonian $H_{eff}$ obtained by
subtracting from jj44b Hamiltonian the isovector P01 or isoscalar P10
interaction (see \cite{Ga62-smdsm,povesplb} for the two-body matrix elements of
$P_{01}$  and $P_{10}$ in $j-j$ coupling).  We have used $G$ =  0.276 for P01
and $G$ = 0.506 for P10 following Refs.  \cite{povesplb,zuker}. In the Figs.
\ref{pairing} (a) and (b) we have shown the contribution of the pairing energies
for $T=0$ odd spin states and $T=1$ even spin states respectively. For the even
$T=1$ levels with $J \le 6^+$, isoscalar pairing plays a larger  role, while for
$T=0$ levels with  $J \le 9^+$, the isovector plays a much greater role. The
$T=1$, $J=0$ ground state has an equal contribution of pairing energies  from
$T=0$ and $T=1$ channels.   The total pairing energy for $T=0$, $J=1$ is 2.14
MeV, whereas for  $T=1$, $J=0$, the total pairing energy is 2.64 MeV. Thus, our
calculation  shows, just as seen in data, that the $T=1$ band should be lower
compared to  the $T=0$ band because of the gain of 0.5 MeV in pairing energy. It
may be noted that our SM calculation  with jj44b interaction predicts the
$T=1$ and $T=0$ band head separation to be  426 keV compared to the experimental
value 836 keV as shown in Fig.  ~\ref{bandsm} and 611 keV from JUN45 interaction
\cite{66as}. 

\begin{figure}
\includegraphics[width=9.5cm]{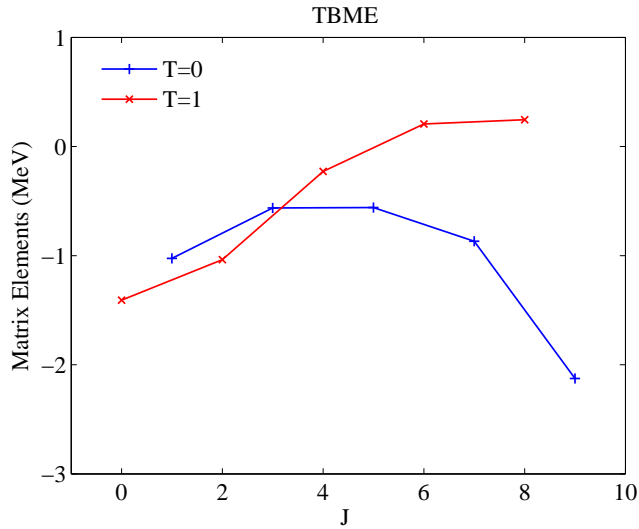}
\caption{Variation of two-body matrix elements $<(g_{9/2})^2 J | V | 
(g_{9/2})^2 J>$ vs $J$.}
\label{smtbme}
\end{figure}

\subsection{Band $\#$4 ($T=0$)}

 The band \#4 (also with $T=0$), as classified in \cite{66as} consists of a level 
with spin $5^+$ at  experimental excitation energy 1.354 MeV  and in addition $9^+$, $11^+$ and
$13^+$ levels; the last three levels are shown in Fig. ~\ref{bandsm}. As we will discuss
ahead, the last three levels form a proper band. 

\subsubsection{$5^+$ and $9^+$ isomeric states}

Firstly, the $5^+$ level in SM  is predicted at excitation energy 0.985 MeV and
DSM predicts the same at 0.893 MeV. However, JUN45 predicts this at 0.407 MeV. 
From DSM it is seen that this level is essentially generated by the oblate
configuration $(3/2^-)_1^{2p,2n} (1/2^-)_1^{2p,2n} (5/2^-\uparrow)_1^{p,n}$ and
the $B(E2)$'s from this level to the $3^+$ levels of bands 2 and band 3 are
relatively small.  Thus, this level is an isomeric state obtained from the
totally aligned $^1f_{5/2}$ $np$ configuration consistent with the claim in
\cite{66as}. Also, the structure of this $5^+$ level is seen to be similar to
the $7^+$ level of band \#3. This is  possibly the reason why in the experiment
reported in \cite{66as}, a large transition strength between these two levels is
seen. The present SM calculation with jj44b interaction gives $B(E2,5^+_1
\rightarrow 3^+_1)$ value to be 45 $e^2fm^4$, DSM gives 1.3 $e^2fm^4$ and the
experimental value \cite{66as} is 13 $e^2fm^4$ (with the effective charges
$e_p=1.5e$ and $0.5e$). Also as given in \cite{66as},
the SM value with JUN45 interaction is 16.02  $e^2fm^4$ (the effective charges
used in Ref. \cite{66as} are $e_p=1.5e$ and $1.1e$). 

Turning to the $9^+$ level of band \#4, in DSM this level (at excitation energy
5.123 MeV and it is much higher than the experimental value)
originates from the oblate intrinsic state with configuration
$(3/2^-)_1^{2p,2n} (1/2^-)_1^{2p,2n} (9/2^+\uparrow)_1^{p,n}$. It has also
strong mixing from several prolate intrinsic states.  The calculated $B(E2)$
values  from this level to the lower $7^+$ levels of bands \#2 and \#3 are
relatively small.  Thus, this level is also predicted to be a isomeric state
with totally aligned $np$ pair in $^1g_{9/2}$ orbit as the dominant structure
and this is consistent with the SM results using JUN45 as  reported in
\cite{66as}.  Experimental value for the $B(E2,9^+_1 \rightarrow 7^+_2)$ is 2.6
$e^2$$fm^4$. The SM values with jj44b interaction is 1.23 $e^2$$fm^4$ and DSM
gives 2.5 $e^2$ $fm^4$. Finally, SM with JUN45 interaction gives the value 0.22 
$e^2fm^4$ \cite{66as}; note that $7^+_2$  level belongs to band \#3.

\subsubsection{$pn$ aligned band with isoscalar $pn$ pair in $g_{9/2}$ orbit}

With the $9^+$ generated by totally aligned $np$ pair in $^1g_{9/2}$ orbit, it
is plausible that the band 4 with $J=9$, $11$ and $13$ can be interpreted as a 
band formed out of coupling the $^{64}$Ge core with a $T=0$ $pn$ pair with the
pair in  $g_{9/2}$ orbit. Keeping this possibility, the $J^\pi=15^+$ and $17^+$
levels are also calculated in SM with jj44b and they are at energies $6.942$ MeV and
$9.285$ MeV respectively.  Note that, as angular momentum is increased, more and
more particles must be put in the $g_{9/2}$ orbital and this will favor a band
at low energy. In Fig.~\ref{smtbme}, shown are the  two-body matrix elements
corresponding to $<g_{9/2}g_{9/2}|V|g_{9/2}g_{9/2}>$.  From the results in the
figure it is clear that the interaction matrix element corresponding to the
$(g_{9/2})^2$ maximally aligned two-particle state $J^\pi$ = $9^+$ (and $T=0$)
is most  attractive (as  reported in ref. \cite{cederwall} -however, the interpretation made in \cite{cederwall} has been heavily criticized
recently in Refs. \cite{cr1,rob,fu,cr3}) although the
$g_{9/2}$ shell is quite high up in energy. 
With $J=9$ aligned pair acting as a
spectator in forming the $T=0$ band in  $^{66}$As with $J$ =9, 11, 13, 15 and
17, the increase of angular momentum simply comes from the core of the remaining
particles. Then, we can describe this band by simply considering  the
ground-state rotational band of the $^{64}$Ge nucleus (the core) with  $J$=0, 2,
4, 6 and 8, and just adding to this band an aligned $pn$-pair providing a
constant energy and a constant angular momentum ($=9$); this will be a isoscalar
pair. 

\begin{figure}
\includegraphics[width=6cm]{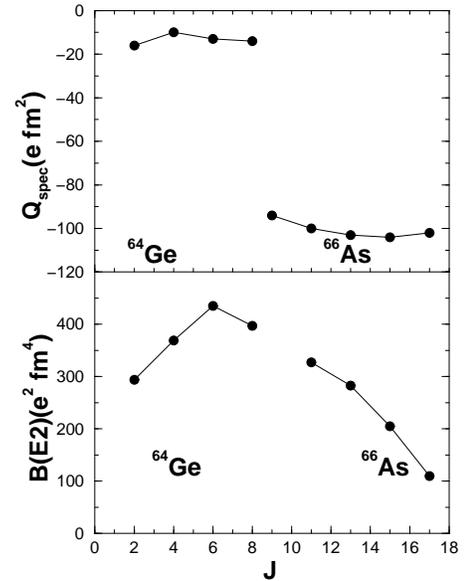}
\caption{Shell model results for $B(E2; J \rightarrow J-2)$ and spectroscopic 
quadrupole  moments for the levels in band \#4 of $^{66}$As  and the ground 
band of $^{64}$Ge.}
\label{be2qq}
\end{figure}

For further understanding of this band and for suggesting experimental
signatures for the structure of this band, $B(E2; J \rightarrow J-2)$ values and
spectroscopic  quadrupole moments of the levels in band \#4 of $^{66}$As and the
levels of the ground band of $^{64}$Ge are calculated in SM and the results are
shown in Fig. \ref{be2qq}. Strikingly, as reported in \cite{pcs(R)} by one of
the authors  (PCS) with the Lund group,  Cranked Nilsson-Strutinsky (CNS)
calculations are quite close to the SM results. In CNS, they are generated by
the rotation axis flips from the intermediate axis to the smallest axis due to
the polarization by the aligned isoscalar $np$ pair in $g_{9/2}$ orbital 
coupled to the $^{64}$Ge triaxial core. Thus, $^{66}$As with band \#4 shows spin
aligned $np$ isoscalar pair phase as seen before in $^{92}$Pd \cite{cederwall}.
As stated in \cite{pcs(R)}, it is important to test the predicted very small
spectroscopic quadrupole moments in $^{64}$Ge and high moments in $^{66}$As.
Let us add that JUN45 predicts $9^+-11^+-13^+$ states at 2506, 3248, 
and 4589 keV but there is no discussion of the structure of these levels
in \cite{66as} where JUN45 results are reported.

\subsection{Band \#5 ($T=0$)}

Ruotsalainen et al \cite{66as} have identified a band of states with 
$J$=$12^{(+)}$, $14^{(+)}$ and $16^{(+)}$ at energies 5.808, 6.530 and 7.792 
MeV. However, they could not assign the parity of these levels uniquely.  SM
calculations were performed for these levels assuming them to be of positive or
negative parity. These levels are shown in Fig. ~\ref{bandsm} as band \#5.  Just
from the energy  systematics, it is more likely that these levels are of
negative parity. This is consistent with the prediction from CNS calculations
(Ragnarsson, private communication).  Experimentally, the collectivity of the
levels in band \#5 is not known. However, SM predictions for the $B(E2)'s$ for
$14^{\pm} \rightarrow 12^{\pm}$ and $16^{\pm} \rightarrow 14^{\pm}$ are $298$
($261$) and $166$ ($332$) $e^2fm^4$, respectively; numbers in the brackets are
assuming negative parity. 

\section{Conclusions}

We have compared the recently available experimental data for $T=0$ and $T=1$
bands for $^{66}$As with the results obtained from shell model 
using jj44b interaction.\\

Following broad conclusions can be drawn:\\

$\bullet$ The present SM calculations with jj44b interaction describe the
observed $T=1$ band (band \#1), isobaric analogue of $^{66}$Ge ground band,
reasonably well and predicted a structural change at  $8^+$. Using DSM it is
seen that at $J=8^+$ there is band crossing originating due to the occupancy of
a proton and a neutron in $g_{9/2}$ orbit.

$\bullet$  We have calculated the pairing energy for $T=0$ and $T=1$ bands
(bands \#3 and  \#1). For the $T=0$ band, the contributions from the isoscalar
pairing is larger for $J\geq9^+$. Also, the $T=1$ band is predicted (as seen in
data)  to be lower  compared to the $T=0$ band because of a gain of 0.5 MeV in
pairing energy.

$\bullet$ The lowest $5^+$ level and the $9^+$ levels are correctly
reproduced by both SM and DSM, with jj44b interaction, to be isomeric states,
as found in the shell model and experimental studies reported in
\cite{66as,Has-2005} with quite small $B(E2)$ values involving these levels.

$\bullet$ The results of shell model for band \#4 in Fig. ~\ref{bandsm} are
described using $^{64}$Ge core coupled to a maximally aligned $pn$ $g_{9/2}$
isoscalar pair (with $T=0$ and $J=9$) giving specific predictions for the
$B(E2)$ values and quadrupole moments for the members of this band. The shell
model results justify a CNS description of this band.

$\bullet$ For band \#5, as the parity of the experimental levels is not known,
shell model predictions for the excitation energies assuming $+ve$ and $-ve$
parity are given in Section III.

$\bullet$  For a better and complete understanding of the structure of bands \#2
and 3 (also \#4), $B(E2)$ values involving the levels of these bands are needed
in future.

$\bullet$ Comparison of the present SM results obtained using jj44b with
those from JUN45 and experimental data showed clearly that jj44b 
is not able to give as good description of the spectroscopy as we have 
before for $^{62}$Ga \cite{Ga62-smdsm}. Similarly, as stated in \cite{66as},
JUN45 also has problems. Thus, clearly there is need to produce
better effective interactions for describing the structure of $N=Z$
nuclei starting from $^{66}$As. 

\acknowledgements

PCS acknowledges the financial support from faculty initiation
grant at IIT Roorkee (India), useful discussions with Profs. S. Aberg
and I. Ragnarsson during  this work and  the hospitality
extended to him during his stay at the Department of Mathematical Physics of the
Lund University. PCS also thanks Profs. T. Mizusaki and N. Shimizu for useful
discussions at  CNS, Tokyo University. R. Sahu is thankful to SERB of Department
of Science and Technology  (Government of India) for financial support.


\begin{thebibliography}{i99}

\bibitem{Sta-07} K. Starosta {\it et al.}, { \color{blue} Phys. Rev. Lett. {\bf 99}, 042503
(2007).}

\bibitem{68se} S.M. Fischer {\it et al.}, { \color{blue} Phys. Rev. Lett. {\bf 84}, 4064
(2000).}

\bibitem{72kr1} G. de Angelis {\it et al.}, { \color{blue} Phys. Lett. B {\bf 415}, 217
  (1997).}

\bibitem{72kr2} H. Iwasaki {\it et al.}, { \color{blue} Phys. Rev. Lett. {\bf 112}, 142502
  (2014).}

\bibitem{72kr3} J. A. Briz {\it et al.}, { \color{blue} Phys. Rev. C {\bf 92}, 054326
  (2015).}

\bibitem{76Sr} A. Lemasson {\it et al.}, { \color{blue} Phys. Rev. C {\bf 85} 041303(R)
  (2012).}

\bibitem{Lister-87} C.J. Lister {\it et al.}, { \color{blue} Phys. Rev. Lett. {\bf 59}, 1270
  (1987).}

\bibitem{cederwall} B. Cederwall {\it et al.}, { \color{blue} Nature {\bf 469}, 68
 (2011).}
 

 
\bibitem{70Br_PRC95} A. I. Morales {\it et al.}, { \color{blue}  Phys. Rev. C {\bf 95}, 064327
  (2017).}
  
\bibitem{Fuller} J. Pruet and G. M. Fuller, { \color{blue} Astrophys. J. Suppl. Ser. {\bf 149},
 189 (2003).}

\bibitem{ks-book} V.K.B. Kota and R. Sahu, Structure of Medium Mass Nuclei: 
Deformed Shell Model and Spin-Isospin Interacting Boson Model (CRC Press, 
Taylor and Francis group, Florida, 2016).

\bibitem{Ga62} H.M. David  {\it et al.},  { \color{blue} Phys. Lett. B {\bf 726},665
 (2013).}
 
 \bibitem{jj44b} B.A. Brown and A.F. Lisetskiy (unpublished); see also endnote
(28) in B. Cheal {\it et al.}, { \color{blue} Phys. Rev. Lett. {\bf 104}, 252502 (2010).}

\bibitem{Ga62-smdsm} P.C. Srivastava, R. Sahu and V.K.B. Kota, { \color{blue} Eur. Phys. J. A
{\bf 51}, 3 (2015).}

\bibitem{66as} P. Ruotsalainen  {\it et al.}, { \color{blue} Phys. Rev. C {\bf 88}, 024320
 (2013).}

\bibitem{sk1} R. Sahu and V.K.B. Kota, { \color{blue} Phys. Rev. {\bf C 66}, 024301 (2002).}

\bibitem{ibm-4} O. Juillet, P. Van Isacker and D.D. Warner, { \color{blue} Phys. Rev. C {\bf 63},
054312 (2001). }

\bibitem{jun45} M. Honma, T. Otsuka, T. Mizusaki and M. Hjorth-Jensen, Phys. 
{ \color{blue} Rev. C {\bf 80}, 064323 (2009).}

\bibitem{No-01}E. Caurier, F. Nowacki, A. Poves and J. Retamosa, Phys.
{ \color{blue} Rev. Lett. {\bf 77}, 1954 (1996).}

\bibitem{rila-2016} R. Sahu and V.K.B. Kota, { \color{blue} Nuclear Theory {\bf 35}, 22 (2016). }

\bibitem{pcs(R)} P.C. Srivastava, S. Aberg and I. Ragnarsson, { \color{blue} Phys. Rev. C 
{\bf 95}, 011303(R) (2017).}



\bibitem{antoine} E. Caurier,  G. Mart\'inez-Pinedo , 
F. Nowacki, A. Poves, and A. P. Zuker, { \color{blue} Rev.\ Mod.\ Phys. {\bf77} (2005) 427.}

\bibitem{nndc} \url{http://www.nndc.bnl.gov/ensdf}

\bibitem{Has-2005} M. Hasegawa, Y. Sun, K. Kaneko and T. Mizusaki,
{ \color{blue} Phys. Lett. B {\bf 617}  (2005) 150-156.}



\bibitem{povesplb} A. Poves and G. Martinez-Pinedo, { \color{blue} Phys. Lett. B {\bf 430}, 
203 (1998).}

\bibitem{zuker} M. Dufour and A.P. Zuker, { \color{blue} Phys. Rev. {\bf C 54}, 1641 (1996).}




 \bibitem{cr1} M. Sambataro and N. Sandulescu, { \color{blue} Phys. Rev. C {\bf 91}, 064318  (2015).}
 
 

 
  \bibitem{rob} S. J. Q. Robinson, T. Hoang, L. Zamick, A. Escuderos, and Y. Y. Sharon, { \color{blue} Phys. Rev. C {\bf 89}, 014316 (2014). }
  
  \bibitem{fu} G. J. Fu, J. J. Shen, Y. M. Zhao, and A. Arima, { \color{blue} Phys. Rev. C {\bf 87}, 044312 (2013).}
  
  

 
  \bibitem{cr3} A. P. Zuker, A. Poves, F. Nowacki, and S. M. Lenzi,{ \color{blue} Phys. Rev. C {\bf 92}, 024320 (2015).}

\end{thebibliography}
\end{document}